\documentclass[nofootinbib,notitlepage,superscriptaddress,10pt,aps,prd,twocolumn]{revtex4-1}
\usepackage[utf8]{inputenc}
\usepackage{amsmath,mathtools}
\usepackage{tensor}
\usepackage{amssymb}
\usepackage{graphicx}
\usepackage{lipsum}  
\usepackage{verbatim}
\newcommand{\leri}[1]{\left(#1\right)}

\begin{document}
\title{Gauge invariant formulation of metric $f(R)$ gravity for gravitational waves}
\author{Fabio Moretti}
\email{fabio.moretti@uniroma1.it}
\affiliation{Physics Department, ``Sapienza'' University of Rome, P.le Aldo Moro 5, 00185 (Roma), Italy}
\author{Flavio Bombacigno}
\email{flavio.bombacigno@uniroma1.it}
\affiliation{Physics Department, ``Sapienza'' University of Rome, P.le Aldo Moro 5, 00185 (Roma), Italy}
\author{Giovanni Montani}
\email{giovanni.montani@enea.it}
\affiliation{Physics Department, ``Sapienza'' University of Rome, P.le Aldo Moro 5, 00185 (Roma), Italy}
\affiliation{ENEA, Fusion and Nuclear Safety Department, C. R. Frascati,
	Via E. Fermi 45, 00044 Frascati (Roma), Italy}
\begin{abstract}
We analyze the propagation of gravitational waves in metric $f(R)$ theories of gravity, on the special setting of flat background geometry (Minkowski spacetime). In particular, adopting a gauge invariant formalism we clearly establish that the exact number of propagating degrees of freedom is three, consisting in the standard tensorial modes along with an additional massive scalar field. Then, investigating their effects on test masses via geodesic deviation equation, we show that the additional dynamical degree contained in such extended formulations is actually detectable as superposition of longitudinal and breathing stresses, which even though in principle corresponding to distinct pure polarizations, turn out to be never separable in the wave dynamics and cannot be interpreted as proper independent excitations. 

\end{abstract}

\maketitle
\section{Introduction}
General Relativity (GR) has a really solid kinematical 
morphology, based on the covariant formulation of the 
spacetime geometry. Nonetheless, its dynamical features 
are more questionable, in view of extensions of the 
Einstein-Hilbert Lagrangian versus more general scalar functions \cite{Bergmann:1968ve,Lovelock:1971yv,Whitt:1984pd,Jackiw:2003pm,Schmidt:2006jt}. 
Among the available choices, when we consider the 
generalization of the Einstein gravity to larger
frameworks, the so-called $f(R)$ models \cite{Cognola:2007zu,Sotiriou:2008rp,Nojiri:2010wj,Berry:2011pb,Capozziello:2019cav} stand for 
their simplicity and viability. Especially, such an extended approach turns out to be very suitable in specific applications, by virtue of its equivalence with Brans-Dicke theories \cite{Bergmann:1968ve,Wagoner:1970vr,Capone:2009xk,ST,Ruf:2017xon}. 
\\ \indent The recent detection of gravitational waves from 
compact objects coalescence \cite{TheLIGOScientific:2016src,Abbott:2017tlp,Abbott:2018utx} suggests that in the near future it 
will be possible to test GR via the 
morphology of the observed gravitational wave template and 
spectra \cite{Maggiore1,Maggiore2,Chatziioannou:2012rf,Maselli:2016ekw,Zhang:2017sym,Blaut:2019fxb,Katsuragawa:2019uto}. Then, it becomes very relevant to be able to predict the modifications induced by extended theory of gravity on the morphology and detectability of spacetime ripples. In this respect, a crucial role is surely played by $f(R)$ models and many efforts have been pursued in past and recent years to suitably characterize the specific track left by this modified theory of gravity. 
\\ \indent  The analysis of linear modes featuring $f(R)$ theories, even on a flat space-time background, suffers from a certain extent of ambiguity concerning the exact number of degrees of freedom carried by the gravitational wave during its propagation. In particular, if in \cite{Capozziello:2008rq} it was claimed the existence of three independent propagation modes, consisting in ordinary tensorial degrees of freedom along with an additional massive scalar mode, other analyses introduced a further massless scalar field, eventually establishing four degrees of freedom \cite{Rizwana:2016qdq}, in contrast with the scalar tensor representation of the theory. A primary contribution of clarification on this literature debate, was provided by the Hamiltonian approach in \cite{Liang:2017ahj}, where the existence of only three independent modes was clearly inferred from the constraint algebra analysis. 
\\ \indent Here, we give a definitive word on this debate, by using fully gauge-invariant quantities to treat gravitational waves and looking the problem both in the scalar-tensor formulation of $f(R)$, as well as in the forth order modified equation approach, obtaining the same firm result: the number degrees of freedom is, as expected by its scalar-tensor formulation, equal to three, i.e. the two standard tensorial modes and an additional scalar one. In this regard, we show also that this extra degree is detectable as a superposition of a longitudinal and breathing stresses. The key point of the obtained results is that these two components, in principle corresponding to distinct pure modes, are actually never separable in the wave dynamics. Eventually, we investigate the nature of the emerging polarizations, in order to get information of the effective morphology that a modified wave could manifest in the interferometers of present and further generations. 
\\ \indent The paper is organized as follows: in Sec.~\ref{Sec2} we study the fourth order equation stemming from metric $f(R)$ theory in terms of gauge invariant variables, outlining the emergence of an additional degree of freedom with respect to GR; in Sec.~\ref{Sec3} we repeat the analysis in the scalar-tensor representation. Eventually in Sec.~\ref{Sec4} conclusions are drawn.

\section{Metric $f(R)$ theories of gravity}\label{Sec2}
The action for a generic $f(R)$ model is given by\footnote{In the following we set $c=1$ and $\kappa=8 \pi G$.}
\begin{equation}
S=\dfrac{1}{2\kappa}\int d^4 x \sqrt{-g} \, f(R) + S_M(g_{\mu\nu},\chi_M),
\label{action f(R)}
\end{equation}
where $f(R)$ is a function of the Ricci scalar $R$ and $S_M$ the action for the matter fields collectively denoted by $\chi_M$, which we assume to be only minimally coupled to the metric. According the metric approach (or second order formulation), we consider the Ricci scalar as a function of the metric variable only, \textit{i.e.}
\begin{equation}
R=g^{\mu\nu}R_{\mu\nu}(g),
\label{eq2}
\end{equation}
with the Ricci tensor obtained by the following contraction of the Riemann tensor
\begin{equation}
R_{\mu\nu}=R\indices{^{\rho}_{\mu\rho\nu}}=\partial_{\rho}\Gamma\indices{^\rho_{\mu\nu}}-\partial_{\nu}\Gamma\indices{^\rho_{\mu\rho}}+\Gamma\indices{^\rho_{\tau\rho}}\Gamma\indices{^\tau_{\mu\nu}}-\Gamma\indices{^\rho_{\tau\nu}}\Gamma\indices{^\tau_{\mu\rho}},
\end{equation}
where the connection components are the usual Christoffel symbols (Levi-Civita connection).
Varying \eqref{action f(R)} with respect to $g_{\mu\nu}$ yields the field equations:
\begin{equation}
    f'(R) R_{\mu\nu}-\dfrac{1}{2}f(R) g_{\mu\nu}-\nabla_{\mu}\nabla_{\nu}f'(R)+g_{\mu\nu}\Box f'(R)= \kappa T_{\mu\nu},
    \label{equation f(R) general}
\end{equation}
where a prime indicates differentiation with respect to the argument and $T_{\mu\nu}$ is the stress energy tensor defined as 
\begin{equation}
T_{\mu\nu}=\dfrac{-2}{\sqrt{-g}}\dfrac{\delta S_M}{\delta g^{\mu\nu}}.
\label{eq4}
\end{equation}
We note that the class of solutions offered by this reformulation is clearly wider than that one viable in GR, as one can easily infer by the inspection of the trace of \eqref{equation f(R) general} (the so-called structural equation), \textit{i.e.}
\begin{equation}
f'(R)R-2f(R)+3\Box f'(R)=\kappa T,\label{f(R)metric_trace}
\end{equation}
that even in the vacuum case ($T=0$) does not compel any more the Ricci scalar to identically vanish, as it occurs in the standard formulation in the absence of cosmological constant.
\\ \indent In the following, we will restrict our attention to metric perturbations around the Minkowski background, i.e.
\begin{equation}
g_{\mu\nu}=\eta_{\mu\nu}+h_{\mu\nu},
\label{pert metr}
\end{equation}
keeping in mind that, in extended theories of gravity, the Minkowski space-time can no longer be considered as the unique "ground state'' of General Relativity (a privileged globally flat state in the quantum dynamics of the gravitational field). Actually, by the inspection of \eqref{f(R)metric_trace}, it is clear that we could deal with equivalently possible "ground states'', as far as we determine forms of the Ricci scalar that are solutions of the vacuum trace equation.
In such cases we could have a non-vanishing background Ricci scalar $R^{(0)}$ and therefore we would be considering gravitational waves propagating on a curved space-time. Although this situation is surely of physical interest, we restrict our analysis to ordinary propagation on Minkowski geometry. Indeed, there exist no experimental evidences that the detection of gravitational waves by present interferometers is affected by an appreciable value of background curvature, except for local turbulent effect, like the Newtonian noise \cite{Fiorucci:2018had}. Our only aim is to clarify the nature of the wave polarization, in view of their possible detection, and to give a contribution to the debate concerning the number of degrees of freedom featuring the considered extended theories. In fact, this information should be not affected by the curvature of the background metric, but it must be an intrinsic information of the gravitational dynamics.
\\ In \eqref{pert metr} we retain $|h_{\mu\nu}|\ll 1$ valid in some reference frame, $\eta_{\mu\nu}=diag(-1,1,1,1)$ and the inverse metric given by
\begin{equation}
g^{\mu\nu}=\eta^{\mu\nu}-h^{\mu\nu},
\label{pert metr inv}
\end{equation}
in order to $g^{\mu\rho}g_{\rho\nu}=\delta^\mu_\nu+\mathcal{O}(h^2)$ be preserved.
At the first order in $h_{\mu\nu}$ the Riemann and the Ricci tensor read as, respectively:
\begin{align}
R^{(1)}_{\rho\sigma\mu\nu}&=\frac{1}{2}\leri{\partial_{\sigma}\partial_{\mu}h_{\rho\nu}+\partial_{\rho}\partial_{\nu}h_{\sigma\mu}-\partial_{\sigma}\partial_{\nu}h_{\rho\mu}-\partial_{\rho}\partial_{\mu}h_{\sigma\nu}}\label{Riemann_lin}\\
R^{(1)}_{\mu\nu}&=\frac{1}{2}\leri{\partial_{\mu}\partial_{\rho}h\indices{^{\rho}_{\nu}}+\partial_{\nu}\partial_{\rho}h\indices{^{\rho}_{\mu}}-\partial_{\mu}\partial_{\nu}h-\Box h_{\mu\nu}}\label{Ricci_ten_lin},
\end{align}
where the trace $h$ is defined as $h\equiv \eta^{\mu\nu}h_{\mu\nu}$ and $\Box\equiv \partial^\mu\partial_\mu$. Lastly, by virtue of \eqref{Ricci_ten_lin}, the Ricci scalar turns out to be
\begin{equation}
R^{(1)}=\eta^{\mu\nu}R^{(1)}_{\mu\nu}=\partial_{\mu}\partial_{\nu}h^{\mu\nu}-\Box h,
\label{Ricci_sc_lin}
\end{equation}
 which is, in general, not vanishing.
Concerning the functional form of $f(\cdot)$, we can imagine to perform a Taylor expansion around the background value $R^{(0)}$, which taken into account \eqref{pert metr} is constrained to be zero. We point out that with respect the analysis pursued in \cite{Olmo:2005hc}, we are considering a spacetime globally flat at the lowest order, neglecting the issues of a non vanishing background curvature on cosmic scale. Accordingly, the function $f(R)$ can be put into the form
\begin{equation}
    f(R)\simeq R+\alpha R^2 + \mathcal{O}(R^3),
    \label{taylorfR}
\end{equation}
that inserted in \eqref{equation f(R) general} carries out
\begin{equation}
    R^{(1)}_{\mu\nu}-\dfrac{1}{2}\eta_{\mu\nu}R^{(1)}-2\alpha\leri{ \partial_\mu \partial_\nu - \eta_{\mu\nu} \Box }R^{(1)}=\kappa T_{\mu\nu}.
    \label{equation f(R) lin}
\end{equation}
Similarly to \eqref{f(R)metric_trace}, by tracing \eqref{equation f(R) lin} we can get a differential equation for the $R^{(1)}$, that is
\begin{equation}
    \left( \Box -m^2\right) R^{(1)}=m^2\kappa T,
    \label{KG R}
\end{equation}
which represents a massive Klein Gordon equation, where $m^{-2}\equiv 6\alpha>0$ holds for $\alpha>0$. Finally, by means of \eqref{KG R} the gravitational field equation can be recast as 
\begin{equation}    R^{(1)}_{\mu\nu}-\dfrac{1}{6m^2}\leri{m^2\eta_{\mu\nu}+2\partial_\mu \partial_\nu }R^{(1)}=\kappa\left( T_{\mu\nu}-\dfrac{1}{3}\eta_{\mu\nu}T\right).  \label{equation f(R) lin 2}
\end{equation}
\\ Following \cite{Weinberg:2008zzc,Flanagan:2005yc}, we introduce for the metric perturbation $h_{\mu\nu}$ the generic decomposition of a rank two symmetric tensor, i.e.
\begin{equation}
\begin{split}
h_{00}&=2\phi ,\\
h_{0i}&=\beta_i+\partial_i \gamma ,\\
h_{ij}&=h^{TT}_{ij} + \dfrac{1}{3}H\delta_{ij}+\partial_{(i}\epsilon_{j)}+\left(\partial_i\partial_j-\dfrac{1}{3}\delta_{ij}\bigtriangleup\right)\lambda,
\end{split}
\label{eq13}
\end{equation}
where $\delta_{ij}$ is the Kronecker delta, $\bigtriangleup\equiv\partial_i\partial^i$ the Laplacian operator and symmetrization is defined as $A_{(ij)}\equiv \frac{1}{2}(A_{ij}+A_{ji})$. 
\\  The irreducible parts introduced in \eqref{eq13} are accompanied by the conditions:
\begin{equation}
\begin{split}
\partial^i \beta_i&=0 \\
\partial^i h^{TT}_{ij}&=0 \\
\eta^{ij}h^{TT}_{ij}&=0\\
\partial^i \epsilon_i &=0,
\end{split}
\label{metric dec}
\end{equation}
which, as stressed in \cite{Flanagan:2005yc} (or more generally in \cite{Weinberg:2008zzc} for a curved background), are required in order to preserve the uniqueness and the consistency of the splitting \eqref{eq13}. Then, it can be demonstrated that under a generic gauge transformation the following combinations of fields, together with $h_{ij}^{TT}$, turn out to be invariant
\begin{equation}
\begin{split}
\Phi &= -\phi + \dot{\gamma}-\dfrac{1}{2}\ddot{\lambda} \\
\Theta &= \dfrac{1}{3} \left ( H-\bigtriangleup \lambda\right)\\
\Xi_i &= \beta_i-\dfrac{1}{2}\dot{\epsilon}_i,
\end{split}
\label{eq15}
\end{equation}
with a dot denoting time derivative. The trace component of the metric perturbation $h_{\mu\nu}$ is now encoded in the gauge invariant scalar fields $\Phi$ and $\Theta$ and, with respect to the usual trace-reverse approach, we are not assuming a priori any condition on $h$. In fact, if in General Relativity the latter
is ultimately non dynamical and can be conveniently made vanishing by a gauge fixing, in $f(R)$ formulations that does not hold anymore and trace contributions cannot be neglected. Indeed, it is easy to show that the first order Ricci scalar can be rewritten as
\begin{equation}
    R^{(1)}=3\ddot{\Theta}-2\bigtriangleup(\Theta+\Phi),
\end{equation}
and by virtue of \eqref{KG R} the trace elements $\Theta$ and $\Phi$ are inherently related to an evolving quantity, and cannot be set to zero. Eventually, the two standard tensorial degrees of freedom for $h_{\mu\nu}$ are instead enclosed in the symmetric transverse and traceless part $h_{ij}^{TT}$, whereas $\Xi_i$ is a divergence free vector (see \eqref{metric dec}), endowed with two independent components. As a result, the decomposition of $h_{\mu\nu}$ in the set of fields $\{\Theta,\Phi,h_{ij}^{TT},\Xi_i\}$ depletes entirely all six of the independent degrees featuring a rank two symmetric tensor for diffeomorphism invariant theory.
\\ Now, in order to unambiguously identify the propagating degrees of freedom, we express the components of $R^{(1)}_{\mu\nu}$ in terms of the set of variables \eqref{eq15}:
\begin{subequations}
  \begin{align}
        R^{(1)}_{00}&=\bigtriangleup\Phi-\dfrac{3}{2}\ddot{\Theta} \\
        R^{(1)}_{0i}&=-\dfrac{1}{2}\bigtriangleup \Xi_i-\partial_i \dot{\Theta} \\
        R^{(1)}_{ij}&= -\partial_{(i}\dot{\Xi}_{j)}-\partial_i\partial_j \left (\Phi+\dfrac{1}{2}\Theta\right)-\dfrac{1}{2}\Box\left(\delta_{ij}\Theta+h_{ij}^{TT}\right).
        \end{align}
\end{subequations}
In turn, a completely analogous decomposition can be performed on the stress energy tensor, namely
\begin{equation}
\begin{split}
T_{00}&=\rho,\\
T_{0i}&=S_i+\partial_i S ,\\
T_{ij}&=\sigma_{ij}+P\delta_{ij}+\partial_{(i}\sigma_{j)} + \left( \partial_i\partial_j-\dfrac{1}{3}\delta_{ij}\bigtriangleup\right)\sigma,
\end{split}
\label{eq16}
\end{equation}
with the relative set of constraints
\begin{equation}
\begin{split}
\partial_i S^i& =0 \\
\partial^i\sigma_{ij}&=0\\
\eta^{ij}\sigma_{ij}&=0\\
\partial_i \sigma^i&=0.
\end{split}
\label{eq17}
\end{equation}
Due to the fact that the stress energy tensor must satisfy the conservation law, that in linearized theory reads as $\partial_\mu T^{\mu\nu}=0$, the irreducible parts just introduced are not independent. Indeed, the following relations must hold
\begin{equation}
\begin{split}
\bigtriangleup S &=\dot{\rho}, \\
\bigtriangleup \sigma &= -\dfrac{3}{2}P+\dfrac{3}{2}\dot{S}, \\
\bigtriangleup \sigma_i&=2 \dot{S}_i.
\end{split}
\label{eq18}
\end{equation}

In terms of the set of variables \eqref{eq15} the field equations \eqref{equation f(R) lin 2} are equivalent to the set of differential equations:
\begin{subequations}
\label{sist}\begin{align}
\bigtriangleup \Phi-\dfrac{3}{2}\ddot{\Theta}+\dfrac{1}{6}R^{(1)}-\dfrac{1}{3m^2} \ddot{R}^{(1)}&=\kappa\left( P+\dfrac{2}{3} \rho \right)\\
 \dot{\Theta}+\dfrac{1}{3m^2}\dot{R}^{(1)} &=-\kappa S \\
 \bigtriangleup \Xi_i &=-2\kappa S_i \\
 \Box h^{TT}_{ij}&=-2\kappa\sigma_{ij}\\
 \dot{\Xi}_i&=-\kappa\sigma_i\\
\Phi+\dfrac{1}{2}\Theta+\dfrac{1}{3m^2} R^{(1)}&=-\kappa\sigma \\
\Box \Theta +\dfrac{1}{3}R^{(1)}&=\dfrac{2}{3}\kappa\left(\bigtriangleup\sigma -\rho \right).
\end{align}
\end{subequations}
Then, by implementing \eqref{KG R} and \eqref{eq18} and operating the substitution $\left(\Phi,\Theta\right)\to \left(\Phi_R,\Theta_R\right)$ with
\begin{equation}
   \Phi_R=\Phi+\dfrac{1}{6m^2}R^{(1)}  \qquad \qquad \Theta_R=\Theta + \dfrac{1}{3m^2} R^{(1)},
   \label{new deg stat f(R)}
\end{equation}
which still provides gauge invariant combinations, we get a much simpler form for system \eqref{sist}:
\begin{subequations}
\begin{align}
\bigtriangleup \Phi_R&=\dfrac{\kappa}{2}\left( 3P+ \rho-3\dot{S} \right)\label{phiR}\\
 \bigtriangleup\Theta_R &=-\kappa \rho\label{thetaR} \\
 \bigtriangleup \Xi_i &=-2\kappa S_i \label{xiR}\\
\label{httR} \Box h^{TT}_{ij}&=-2\kappa\sigma_{ij}.
\end{align}
\end{subequations}
By the inspection of the system \eqref{phiR}-\eqref{httR} we see that the two tensorial modes $h_{ij}^{TT}$ are the only propagating degrees of freedom, in conjunction with $R^{(1)}$. The two modified scalars $\Phi_R$ and $\Theta_R$ together with the vector $\Xi_i$ are instead solutions of Laplace equation and they cannot exhibit radiative behaviour. Therefore, we claim that metric $f(R)$ theories are naturally equipped with three independent propagating degrees, in agreement with the results of \cite{Liang:2017ahj}. 
\\Now, since we are interested in analyzing the effects due to \eqref{KG R} and \eqref{httR} on a sphere of test masses, it may be instructive to consider the geodesic deviation equation in the comoving frame, having set $T_{\mu\nu}=0$. In this case, the only relevant Riemann components are given, in terms of gauge invariant quantities, by
\begin{equation}
    R_{i0j0}=-\dfrac{1}{2}\ddot{h}_{ij}^{TT}+\partial_{(i}\dot{\Xi}_{j)}+\partial_i\partial_j\Phi-\dfrac{1}{2}\delta_{ij}\ddot{\Theta}.
    \label{geod f(R)}
\end{equation}
Even if it seems that the only proper dynamical effects be induced by the ordinary $h_{ij}^{TT}$ gauge invariant part, we stress the fact that now the truly static degrees are not $\Theta$ and $\Phi$. In fact, by virtue of \eqref{new deg stat f(R)} they actually depend on the propagating degree $R^{(1)}$ and, solving these relations for the new static components $\Theta_R$ and $\Phi_R$, expression \eqref{geod f(R)} can be recast as
\begin{multline}
    R_{i0j0}=\overbrace{\partial_{(i}\dot{\Xi}_{j)}+\partial_i\partial_j\Phi_R-\dfrac{1}{2}\delta_{ij}\ddot{\Theta}_R}^{\text{Static part}}+ \\
   +\underbrace{\alpha \leri{\delta_{ij}\partial_t^2-\partial_i\partial_j }R^{(1)} -\dfrac{1}{2}\ddot{h}_{ij}^{TT}}_{\text{Radiative part}}.
\end{multline}
Now, if we consider a gravitational wave travelling along the $z$ axes, the contributes of the solely $R^{(1)}$ mode to the geodesic deviation equation are given by
\begin{equation}
\begin{split}
&\frac{\partial^2 \delta_x}{\partial t^2}\simeq\frac{R^{(1)}}{6}\leri{1+\frac{k_z^2}{m^2}}\;x_0\\
&\frac{\partial^2 \delta_y}{\partial t^2}\simeq\frac{R^{(1)}}{6}\leri{1+\frac{k_z^2}{m^2}}\;y_0\\
&\frac{\partial^2 \delta_z}{\partial t^2}\simeq\frac{R^{(1)}}{6}\;z_0
\end{split}
\label{geodesic R}
\end{equation}
where we set the vector denoting the separation between two nearby geodesics as
\begin{equation}
    \vec{x}=(x_0+\delta_x,y_0+\delta_y,z_0+\delta_z),
    \label{vec}
\end{equation}
with $x_0$ and $\delta_x$ indicating the rest position and the displacement of order $\mathcal{O}(h)$ induced by the wave, respectively\footnote{Analogously for $y,z$.}. Lastly, the wave vector for $R^{(1)}$ is fixed in $k^\mu=(\sqrt{k_z^2+m^2},0,0,k_z)$.
\\ We remark that the phenomenology associated to the $R^{(1)}$ mode is actually the result of a superposition of two distinct kind of polarizations. In particular, from \eqref{geodesic R} is easy to identify in $R^{(1)}$ excitations a breathing mode acting on the transverse plane $xy$ and a longitudinal mode along the $z$ direction, which is moreover independent on the mass parameter $m$. In this regard, it is useful to define the polarization matrices\footnote{We point out that such a decomposition is strictly applicable only in the case of waves following null geodesics, i.e. in the presence of massless modes. Therefore, despite the application of this method is not formally allowed in this case, we suggest it could still give a precious physical insight about the emerging new phenomenology.} $\mathbb{E}_b$ and $\mathbb{E}_l$ (see \cite{Isi:2015cva} for a comparison):
\begin{equation}
\mathbb{E}_b\equiv
\begin{pmatrix}
1 & 0 & 0 \\
0 & 1 & 0 \\
0 & 0 & 0
\end{pmatrix}\qquad\mathbb{E}_l\equiv
\begin{pmatrix}
0 & 0 & 0 \\
0 & 0 & 0 \\
0 & 0 & 1
\end{pmatrix},
\end{equation}
which allows us to recast \eqref{geodesic R} as
\begin{equation}
  \partial_t^2\delta\vec{x}=\leri{P_1\,\mathbb{E}_b+P_2\,\mathbb{E}_l}\vec{x}_0
  \label{geod matr1}
\end{equation}
with $P_1\equiv \frac{R^{(1)}}{6}\leri{1+\frac{k_z^2}{m^2}}$, $P_2\equiv \frac{R^{(1)}}{6}$ and $\vec{\delta}_x$, $\vec{x}_0$ representing the rest and the $\mathcal{O}(h)$ component of \eqref{vec}, respectively.
However, we can introduce a new set of polarization matrices $\{\mathbb{E}_t,\,\mathbb{E}_d\}$, related to the irreducible representations for a spin 2 particle described by a symmetric tensor of rank 2, namely
\begin{equation}
\mathbb{E}_t\equiv \frac{1}{3}
\begin{pmatrix}
1 & 0 & 0 \\
0 & 1 & 0 \\
0 & 0 & 1
\end{pmatrix},\quad
\mathbb{E}_d\equiv \frac{2}{3}
\begin{pmatrix}
1/2 & 0 & 0 \\
0 & 1/2 & 0 \\
0 & 0 & -1
\end{pmatrix},
\end{equation}
where $\mathbb{E}_t$ is the trace part concerning the state $|0,0\rangle$ and $\mathbb{E}_d$ the traceless component pertaining the state $|2,0\rangle$.
\\ Then, expressing the set $\{\mathbb{E}_b,\,\mathbb{E}_l\}$ in terms of $\{\mathbb{E}_t,\,\mathbb{E}_d\}$ 
the geodesic deviation \eqref{geod matr1} can be rearranged as
\begin{equation}
  \partial_t^2\delta\vec{x}=\frac{1}{3}\leri{(2P_1+P_2)\,\mathbb{E}_t+(P_1-P_2)\,\mathbb{E}_d}\vec{x}_0.
    \label{geod matr2}
\end{equation}
Therefore, we see from \eqref{geod matr2} that the additional degree of freedom $R^{(1)}$ is actually capable to excite two distinguished scalar parts of the available modes for the $h_{\mu\nu}$ tensor, and this is ultimately due to the fact that trace free condition cannot be achieved in $f(R)$ gravity in Minkowski background as well. Such an outcome is also consistent with the fact that, even in the absence of ordinary matter, $f(R)$ theories are equipped with an effective stress energy tensor related to the additional degree of freedom, as it can be appreciated in the scalar-tensor reformulation (see Sec.~\ref{Sec3}). In this respect, we note that it is in agreement with ordinary General Relativity results, where the trace of $h_{\mu\nu}$ cannot be set vanishing within matter sources.

\section{Scalar-tensor representation for $f(R)$ models}\label{Sec3}
Scalar-tensor representations of $f(R)$ theories represent a useful tool for investigating the enlarged dynamical content of these models. In particular, such a reformulation enables us to single out the additional degree of freedom contained in the functional form $f(\cdot)$, by means of a scalar field non minimally coupled to gravity and self-interacting. Hence, let us rearrange \eqref{action f(R)} as
\begin{equation}
S=\frac{1}{2\kappa}\int d^4x \sqrt{-g}\leri{f(\Upsilon)+f'(\Upsilon)(R-\Upsilon)}+S_M,\label{f(R)metric_xi}
\end{equation}
where $\Upsilon$ is an auxiliary field whose variation carries out the dynamical condition $\Upsilon=R$, provided $f''(R)\neq 0$. Then, introducing the scalar field $\xi\equiv f'(R)$, action \eqref{f(R)metric_xi} can be recast in the scalar-tensor form\footnote{Especially, expression \eqref{f(R)metric_xi} can be considered as a Brans-Dicke model of parameter $\omega=0$.}
\begin{equation}
S=\frac{1}{2\kappa}\int d^4x \sqrt{-g}\leri{\xi R - V(\xi)}+S_M,\label{f(R)metric_phi}
\end{equation}
the potential term $V(\xi)$ being defined by
\begin{equation}
V(\xi)\equiv \Upsilon(\xi)\xi-f(\Upsilon(\xi)).
\end{equation}
Variation of \eqref{f(R)metric_phi} with respect to the metric and the field $\xi$ gives us, respectively:
\begin{equation}
    G_{\mu\nu}-\frac{1}{\xi}\leri{\nabla_\mu\nabla_\nu-g_{\mu\nu}\Box}\xi+\frac{1}{2\xi}g_{\mu\nu}V(\xi)=\frac{\kappa}{\xi}T_{\mu\nu}
    \label{STmetric_eqg}
\end{equation}
and
\begin{equation}
    R=V'(\xi),
    \label{STmetric_eqphi}
\end{equation}
where $G_{\mu\nu}=R_{\mu\nu}-\frac{1}{2}g_{\mu\nu}R$. Eventually, by close analogy with \eqref{f(R)metric_trace}, taking the trace of \eqref{STmetric_eqg} and using \eqref{STmetric_eqphi}, it is easy to get a dynamical equation for the scalar field $\xi$, that is
\begin{equation}
3\Box\xi+2V(\xi)-\xi V'(\xi)=\kappa T\label{STmetric_trace}.
\end{equation}
Now, since we are interested in the linearized limit of the theory (see \eqref{pert metr}), it is reasonable to assume also for the field $\xi$ the perturbative expansion
\begin{equation}
    \xi=\xi_0+\delta\xi,
    \label{pert xi}
\end{equation}
where $\delta\xi=\mathcal{O}(h)$. Thus, by virtue of \eqref{pert xi}, we can expand the potential $V(\xi)$ around $\xi_0$ as well, \textit{i.e.}
\begin{equation}
    V(\xi)\simeq V(\xi_0)+V'(\xi_0)\delta\xi+\frac{1}{2}V''(\xi_0)\delta\xi^2.
    \label{pert V}
\end{equation}
However, when  we required Minkowski background be a solution, i.e. $R^{(0)}=0$, from \eqref{STmetric_eqphi} it follows the constraint $V'(\xi_0)=0$, and consistency at the lowest order for \eqref{STmetric_trace} leads to the additional condition $V(\xi_0)=0$. That also guarantees that in \eqref{STmetric_eqg} cosmological constant contributions do not appear, in which case we should instead deal with de Sitter-like backgrounds. Of course, the requirement of having a stable minimum in $\xi_0$ for the potential $V(\xi)$ cannot be always attainable \cite{Clifton:2006ug,Nojiri:2017ygt}, but in the following we shall disregard these cases. 
\\Now, analogously to \eqref{KG R} we can rearrange \eqref{STmetric_trace} in the following way
\begin{equation}
    \leri{\Box-M^2}\delta\xi=\frac{\kappa}{3} T,
    \label{KG xi}
\end{equation}
where we define $M^2=\frac{\xi_0 V''(\xi_0)}{3}$ and we require $V''(\xi_0)>0$ in order to avoid instabilities of the solution.

We write \eqref{STmetric_eqg} at first order in $h_{\mu\nu}$ and $\delta\xi$, obtaining
\begin{equation}
    R^{(1)}_{\mu\nu}-\dfrac{1}{2}\eta_{\mu\nu}R^{(1)}-\leri{\partial_\mu\partial_\nu - \eta_{\mu\nu}\Box}\zeta=\kappa ' T_{\mu\nu} ,
\end{equation}
where we have introduced $\kappa'\equiv \kappa / \xi_0$ and $\zeta\equiv \delta\xi/\xi_0$.
Accordingly with the previous section, we obtain a set of ten differential equations that must be solved together with equation \eqref{KG xi}. Therefore, we remove the redundancies by virtue of \eqref{eq18} and we define the new couple of scalars $\leri{\Phi_\zeta,\Theta_\zeta}$ in the following way
\begin{equation}
    \Phi_\zeta=\Phi+\dfrac{1}{2}\zeta \qquad  \qquad \Theta_\zeta=\Theta+\zeta.
\end{equation}
Lastly, we manage to reduce the number of independent equations to six:
\begin{subequations}
	\begin{align}
	\bigtriangleup \Phi_\zeta&=-\dfrac{1}{2}\kappa' \left(3\dot{S}-3P-\rho\right)\\
	\bigtriangleup \Theta_\zeta &=-\kappa' \rho \\
	\bigtriangleup \Xi_i &=-2\kappa'S_i \\
	\Box h^{TT}_{ij}&=-2\kappa'\sigma_{ij}.
	\end{align}
\end{subequations}
As in the metric case it is easy to recognize that only the tensorial degrees $h_{ij}^{TT}$ and the massive scalar $\zeta$ are solutions of wave-like equations, whilst the transverse vector $\Xi_i$ and the couple of scalars $\leri{\Phi_\zeta,\Theta_\zeta}$ are static.
Given the fact that the two transformations $\leri{\Phi,\Theta} \to \leri{\Phi_R,\Theta_R}$ and $\leri{\Phi,\Theta} \to \leri{\Phi_\zeta,\Theta_\zeta}$ are identical if one makes the identification $2\alpha R^{(1)}\leftrightarrow\zeta$, it follows that the phenomenology associated with the presence of the massive scalar $\zeta$ is the same described in the previous section. Especially, the amplitudes of the polarizations are rescaled by the $\xi_0$ factor and the coincidence of the results is reached if $\xi_0\to 1$, corresponding to \eqref{taylorfR}, where the coefficient of the linear term is set to unity.

\section{Concluding Remarks}\label{Sec4}
We investigated the nature of the oscillating linear modes that the gravitational field outlines in metric $f(R)$ models. In particular, our analysis relies on a gauge-invariant formulation of the gravitational wave propagation, obtained from a linear expansion of the dynamics near the Minkowski spacetime.
\\ \indent The necessity for such an approach finds its justification
in the debate present in literature, concerning the number of degrees of freedom that the theory possesses in the considered formulation. In this regard, the advantage of a gauge-invariant study consists in the possibility to unambiguously identify the actual propagation modes, in order to evaluate their
effects on test particles arrays via the geodesic deviation.
\\ \indent We clarified out of any doubts that only three independent dynamical degrees are present in the linear theory, consistently with the degrees of freedom of the full non-linear theory.  Our findings are also compatible with \cite{Suvorov:2019qow}, where perturbations around a Kerr metric are studied.\\
\indent We elucidated that the source of the debate regarding the exact number of modes is originated by a very subtle feature of the obtained gauge invariant formulation. Indeed, even though we deal with a single scalar degree
in addition to the metric field of the standard Einstein-Hilbert action, such a  degree appears in the geodetic deviation equation like the superposition of two different and well known polarizations, i.e. a breathing and a longitudinal mode.
However, we remarked that these two polarizations can be never physically separated and they always act on particle arrays like a single one, bringing features of its basic constituent at the same level.
\\ \indent  We stress that a characteristic feature of General Relativity is the traceless nature of vacuum gravitational waves. Especially, in Minkowski space-time this property is an exact output of the dynamics, while on a curved background it is valid up to higher order terms \cite{Maggiore1,Misner:1974qy,Straumann:2013spu}. In $f(R)$ theories of gravity, instead, the traceless character of vacuum gravitational waves is intrinsically lost both in Minkowski and curved backgrounds. Indeed, our analysis clearly elucidates the dynamical and non-vanishing nature of the trace and this feature is expected to be preserved on any background, since the gauge invariant variables method can be applied straightforwardly to the general case (see \cite{Weinberg:2008zzc}). The reason for such a different morphology of modified gravitational waves consists of the scalar-tensor nature of the $f(R)$ model, in which the non-minimally coupled scalar field plays the role of matter source. With this regard, we suggest that the excitation of the trace part of the metric tensor could be due to the presence of an effective stress energy tensor, related to the additional scalar degree. Exactly as in General Relativity the passage from flat to curved space-time does not alter the number of degrees of freedom associated to the wave, at least at the considered order of approximation, analogously we expect that the two tensorial and the scalar modes of the present Minkowski model unaltered survive in the same curved space-time extension. This firm conjecture is also validated by the Hamiltonian analysis performed in \cite{Liang:2017ahj}, verifying the first class nature of the emerging Hamiltonian constraints.
\\  \indent It has been demonstrated in \cite{Kijowski:2016qbc,Afonso:2018hyj} that a class of extended theories of gravity, which $f(R)$ models belong, is equivalent to standard General Relativity possibly coupled to scalar fields, which may not coincide with the scalar field we deal with in the Jordan frame. This interpretation is certainly viable, and without entering the details of a possible quarrel about the geometrical or matter representation of this more general framework, we observe that the present analysis about the number of physical degrees of freedom remains significant in both these scenarios, even if the associated phenomenology could be altered by the peculiar representation adopted. \\ \indent In conclusion, since the morphology of the theory is now under control and the actual features of the gravitational polarizations are well-traced, it is possible to better characterize the phenomenological implications and the detectability of such extended gravitational ripples.

\end{document}